# Optimal Clustering of Energy Consumers based on Entropy of the Correlation Matrix between Clusters

Nameer Al Khafaf, Mahdi Jalili and Peter Sokolowski
Electrical and Biomedical Engineering
RMIT University, Melbourne, Australia

*Abstract*—Increased deployment of residential smart meters has made it possible to record energy consumption data on short intervals. These data, if used efficiently, carry valuable information for managing power demand and increasing energy consumption efficiency. However, analyzing smart meter data of millions of customers in a timely manner is quite challenging. An efficient way to analyze these data is to first identify clusters of customers, and then focus on analyzing these clusters. Deciding on the optimal number of clusters is a challenging task. In this manuscript, we propose a metric to efficiently find the optimal number of clusters. A genetic algorithm based feature selection is used to reduce the number of features, which are then fed into self-organizing maps for clustering. We apply the proposed clustering technique on two electricity consumption datasets from Victoria, Australia and Ireland. The numerical simulations reveal effectiveness of the proposed method in finding the optimal clusters.

## I. INTRODUCTION

With the advancement in energy metering technologies and infrastructure, energy consumers now have access to data recorded by smart meters on regular intervals, usually every 15 or 30 minutes. These data, in its raw format, hold little or no value to electricity market stakeholders. However, by using the right artificial intelligence and machine learning techniques, the raw data can be transformed into valuable data. [1] proposed a system based on machine learning approach to transform raw data recorded from the meters into consumption predictions as well as recommendation and for energy users to improve energy efficiency. Advantage of such a system is that it uses existing metering infrastructure to improve energy efficiency without the need for installing new hardware. [2] proposed a hybrid information and communication technology solution to provide analytics for energy users in a form of web portal application, where energy users can use it to find insights and better understand their bills and consumption rates. This shows the potential advantages of smart meter's data in energy efficiency if the right tools are used [3]. Other capabilities for metering data include load forecasting [4], pricing intelligence [5] and demand response programs [6, 7].

There are several challenges associated with the data generated by smart meters. Based on 30-minute energy reading, the smart meters record 17,520 data points per year for a single customer. With millions of energy consumers, the number is easily translated to billions of data points, making it a difficult task to efficiency mine the data [8, 9]. Other challenges include the speed for processing and gaining knowledge from meter's data as they are recorded, where sometime real-time actions are required to make. This calls for advanced analytics solutions to transform these data into knowledge and value-added products [10]. A number of existing works have already tried to address the challenge. Bedingfield et al. [11] addressed key challenges of processing meter's data and proposed a scalable algorithm for data processing, which has been tested on 10,000 householders for a year in the Australian electricity market. Other form of addressing the challenges to meter's data is through knowledge discovery by segmenting energy consumers based on daily energy consumption recorded by smart meters [12].

Consumer segmentation comes in different forms referred to as clustering, that is to group consumers into clusters based on similarity in energy consumption. Haben et al. [13] analyzed consumers' demand and found distinct behavior in consumption associated with each group by using a finite mixture of Gaussian distribution for clustering. [14] proposed a clustering method based on the Euclidean distance between load patterns of different consumer. They showed that clustering techniques are effective in identifying outliers in a group of energy consumers. Other methods of load profile clustering include multi-resolution clustering in spectral domain [15], which was shown to be superior to traditional K-means clustering.

One distinct method for clustering of energy consumption is based on the use of Self-Organizing Maps (SOMs). Verdu et al. [16] proposed the use of SOMs to identify electrical load patterns for customers. The results showed that SOMs are suitable for identifying and classifying electrical users into clusters. Sanchez et al. [17] used SOMs to segment clients based on their domestic energy consumption. However, both these works stopped at clustering customers into pre-determined number of groups rather than finding the optimal number of clusters for a given dataset. Therefore, the main aim of this research is to extract knowledge from the raw information recorded by smart meters by identifying the optimal number of clusters for a give dataset.

As we have a huge number of features, we first use a feature selection method to reduce the number of features. Often, in many situations, machine learning tasks such as clustering or classification requires using many features, and it is well-known that generalizability of models decreases by increasing the feature space. A number of methods have been proposed to reduce the feature space, and feature selection has been shown to be an effective way of doing so [18, 19]. Feature selection is the process of searching for the most significant features for a predictive model to save in computational cost and eliminate unneeded features that could affect the performance of the predictive model [20]. Feature selection can indeed be modeled as an optimization problem, where population-based optimization methods can be effectively used to solve it. For example, Genetic Algorithms (GA) has been proposed for the optimal feature selection problem [21]. Here we use GA to select the most informative feature for the clustering purpose. The reduced features are then given to SOM for clustering.

We use an innovative approach to decide the optimal number of clusters. The novelty behind the proposed metric is to maximize distinguishability between the clusters. Our proposed metric is a simple metric based on the eigenvalues of the correlation matrix between the clusters. The proposed optimal clustering method is applied on two datasets containing 609 residential customers in Victoria, Australia and 4,225 residential customers in Ireland. Our numerical results show that the proposed technique is effective is correctly identifying the optimal clusters.

## II. Data Collection and Pre-processing

### A. Dataset 1: Australian Energy Consumption

The energy consumption profiles of 609 households in Victoria, Australia were recorded anonymously for a full year by a Power Distribution Network Authority using residential smart meters. The smart meter records a net consumption/generation on a second by second basis, which aggregates the net consumption into half an hour of the entire individual usage block by the second. In other word, the meter could measure consumption in the first second and generation in the second the customer has solar energy installed at the premise. The energy consumption profile comprised of 48 data points across the day averaging over 30 minutes. The data points were collected over duration of one year recorded in 9 distinct files where each file represents a time period in the year as depicted in Table 1. This represents the raw data obtained by smart meter.

Most of the existing works in the literature have used average energy consumption over a period of time; a week, a month or a season, in their model. However, in doing that many of the features are often lost, especially in a large datasets where high consumption cancels out low consumptions. To this extent, we used the whole one-year dataset in our model for all consumers. The raw data comprised of a column vector 10,698,912-by-1 for the whole dataset of 609 consumers over the year. A data pre-processing step transformed the column vector into 17,498-by-609 matrix where rows represent time periods, starting from March 2015 and ending in March 2016, and columns represent energy consumers. We removed 42,630 data point from the original raw column vector to maintain the same time period for all consumers as some of the smart meters recorded either earlier or later than others. Figure 1 summarizes the raw and pre-processed data.

Figure 2 depicts energy consumption profiles of a sample of four energy consumers in the feature matrix for the whole year. This feature vector is an input to the genetic algorithm in the following section.

TABLE I
SMART METER RAW DATA PROPERTIES

| Data File No. | Start Date | End Date | No. of Sample Points per Consumer |
|---|---|---|---|
| 9 | 9-Mar-15 | 19-Apr-15 | 2,016 |
| 1 | 20-Apr-15 | 31-May-15 | 2,016 |
| 2 | 1-Jun-15 | 11-Jul-15 | 1,968 |
| 3 | 12-Jul-15 | 22-Aug-15 | 2,016 |
| 4 | 23-Aug-15 | 2-Oct-15 | 1,968 |
| 5 | 3-Oct-15 | 14-Nov-15 | 2,064 |
| 6 | 15-Nov-15 | 27-Dec-15 | 2,064 |
| 8 | 28-Dec-15 | 6-Feb-16 | 1,968 |
| 7 | 7-Feb-16 | 8-Mar-15 | 1,488 |

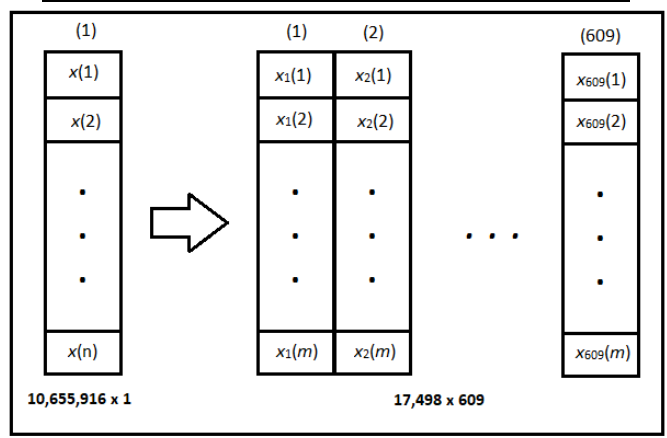

Fig. 1. Raw and pre-processed data of consumption profile of 609 energy consumers obtained from an energy distribution company in Victoria, Australia.

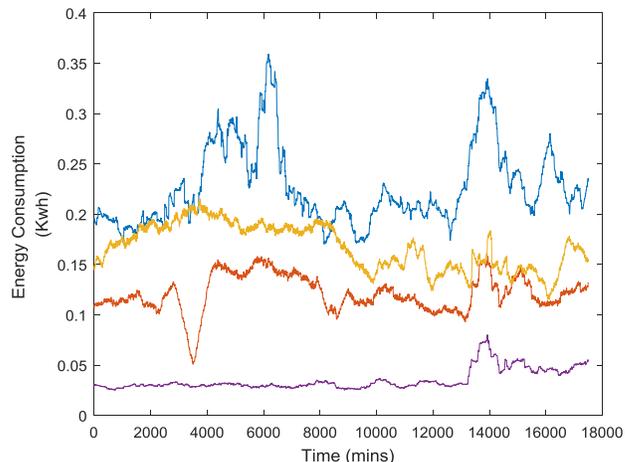

Fig. 2. Energy consumption profiles of four energy consumer in the dataset selected at random. Figure shows unique consumption pattern across the year.

*B. Dataset 2: Irish Energy Consumption*

The Commission for Energy Regulation in Ireland conducted an electricity behavioral trial on Irish homes and businesses participating in the trial during 2009 and 2010. The purpose of the trial was to assess the cost-benefit of smart meters national roll out and collection of energy consumption to support the assessment [26]. Further detail on the study and the findings can be found in [27]. The outcome of the trial is a total of 6,445 energy consumption out of which 4,225 were residential, 485 were small to medium enterprises and 1,735 were other users that did not fall in the first two categories. Since this research focuses on residential energy consumption, only 4,225 energy consumers were selected for optimal clustering.

The residential energy consumptions were recorded every 30 minutes starting in mid-June 2009 to late December 2010 resulting in a column vector 108,713,475-by-1 for the whole dataset. Following a similar approach utilized on dataset 1, the raw data was rearranged into a feature matrix 25,731-by-4,225 where rows represent the energy consumption and columns represent the energy consumer. The feature matrix was further reduced into 17,498-by-4,225 to match with the one-year energy consumption of dataset 1. The feature matrix and the energy consumption profile are similar to those depicted in Figures 1 and 2, respectively.

### III. Feature Selection and Clustering

*A. Genetic Algorithm (GA)*

Given an energy consumption feature matrix of size *M*-by-*N*, where *M* is the number of features per energy consumer and *N* is the number of energy consumers, an initial population matrix of size *K*-by-*V* is created, where *K* is the population size and *V* is number of design variables. Each row in the feature matrix is a unique time instance for all energy consumers and each row in the population matrix is a random sample of row indices of time instances of the feature matrix for the initial population.

The GA optimizer takes on three inputs: a fitness (or objective) function, the number of design variables and the settings to return the index for the most significant selected features in the feature vector. The fitness function follows a standard rank-based selection strategy using a linear combination of the error rate, in which individuals are ranked based on the error rate and the fitness value for each individual is assigned based on the rank of the individual. The fitness function requires the feature matrix and the class labels vector. The settings for GA include the population size, the number of generations, the probability of crossover and that of mutation. The crossover operator combines selected individuals with unique features to generate new individuals with features from both parents. The mutation operator changes one feature in the new individual at random. These guarantee a variety combination of features in a population. The output of the feature selection GA feeds into the clustering model in the next section.

*B. Self-Organizing Maps (SOMs)*

SOM is a type of unsupervised artificial neural network introduced by Kohonen [22], and is often used for dimensionality reduction such as clustering based on a similarity concept. Using SOMs for customer clustering, energy consumption segmentation is performed by grouping energy consumers based on the energy consumption across the year where each time period is a unique instance. SOMs use a competitive layer to group a dataset feature matrix of any dimensions to as many clusters as the neuron of the competitive layer. The clustering depends on the Euclidean distance between the input vector and the weight vector of all neurons. A neuron is called the best matching unit if its weight vector is most similar to the input. The weight of the best matching unit and all other closer neurons are adjusted toward the input vector. Given a feature vector *F*, neuron *i* updates its weight vector $W_i$ as:

$$W_i(s+1) = W_i(s) + a(F(t) - W_i(s)) \qquad (1)$$

where *s* is the iteration step, *a* is the learning coefficient and *t* is the feature vector index.

An SOM is created with a pre-defined number of clusters based on the number of neurons and is trained with the optimal set of features obtained through GA. The initialized SOM uses the feature vector as an input and updates the weight of each neuron toward the input vector to group the data into clusters. For a better clustering, one can use additional customers' information such as location, age, number of people in the house. However, due to the anonymity of such data, only the energy consumption is available as an input. SOM produces a binary cluster matrix *A*-by-*B*, where *A* is the number of clusters and *B* is the number of energy consumers. The binary element of the cluster matrix indicates which consumer is within which cluster; where 1 indicates presence of the consumer in a cluster and 0 indicates absence of the consumer from the cluster.

To discover a representative energy consumption profile for each cluster, the energy consumptions of all energy consumers within a cluster are averaged over the same time period across the year. This produces cluster vector of 17,498-by-1 for each cluster. Due to high variability of energy consumption over the year, the plot of the cluster vector is difficult to interpret and hence a moving average filter of window of 700 samples is applied to smoothen the cluster's representative energy consumption profile. Figure 3 shows a cluster vector across the year before and after applying the moving average filter. It can be observed that both time series follow similar patterns, but the one in red provides a better visualization of the energy consumption.

To find the optimal number of clusters for a given dataset, we introduce an optimal cluster metric *C*. It is based on entropy of the eigenvalues of the correlation matrix of the clusters [23, 24]. The originality behind this metric is that clusters should have as less similar information as possible. In other words, they should have maximum distinguishability. Let *L* be the correlation matrix *A*-by-*A* where *L* is a symmetric matrix and each entry of the matrix is the correlation coefficient between clusters. Our proposed metric to measure quality of clustering is defined as

$$C = 1 + \frac{\sum_{i=1}^{A} \lambda_i' \log(\lambda_i')}{\log(A)}; \lambda_i' = \frac{\lambda_i}{\sum_{i=1}^{A} \lambda_i} \quad (2)$$

where $A$ is the number of clusters defined previously and $\lambda_i$ is the $i$-th eigenvalue of $L$. $C$ ranges between 0 indicating uncorrelated (maximum distinguishability) between clusters, and 1 indicating perfectly correlated (minimum distinguishability) between clusters. In order to find the optimal number of clusters, one can vary the number of clusters and observe the value of $C$. The case with the lowest $C$ is the optimal one.

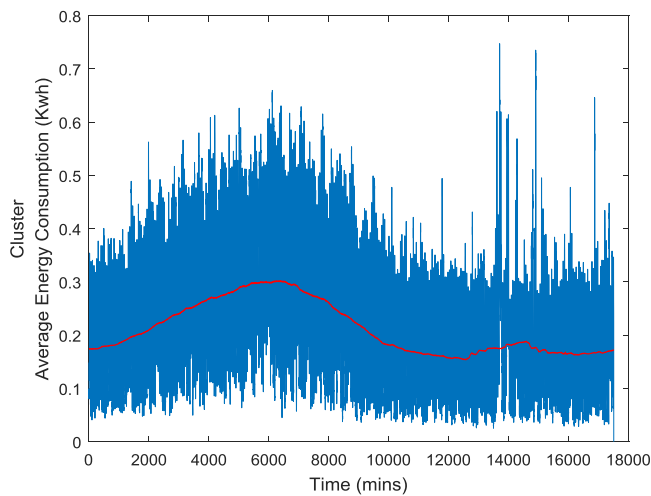

Fig. 3. Cluster's average energy consumption (in blue) and moving average of cluster's average energy consumption (in red). The cluster's energy profile is noisy and it is difficult to interpret without smoothing the signal.

IV. RESULTS AND DISCUSSIONS

A. Dataset 1

Using GA-based feature selection method proposed in section II, the settings for the algorithm were determined based on preliminary number of runs as follows: Population size: 200, Number of generations: 120, Probability of crossover: 0.3, Probability of mutation: 0.001. Table 2 provides 12 times instances where GA flagged as significant features. Figure 4 illustrate selected features at two time instances; 22-Sep-2015 21:30:00 and 11-Mar-2015 18:00:00. It can be observed from the figure that there are many energy consumers with energy consumptions more than 1 Kwh at that time instance which shows the effectiveness of the GA in identifying peaks energy consumptions.

In order to find the optimal number of clusters for these 609 energy consumption profiles, SOM is used to cluster the data into clusters by varying the cluster size from 3 to larger values, and observing $C$ for each case. At each iteration, we initialize SOM to cluster the energy consumption profiles, deduce the cluster average energy consumption profile and calculate $C$. The process stops when $C$ does not decrease from the lowest recorded value for 5 consecutive iterations. Figure 5 shows how $C$ changes by varying the cluster size from 3 to 9. A cluster size of 2 was not considered since much information will be lost in averaging the dataset over two clusters only. By increasing the cluster size from 3, $C$ declines indicating that the quality of the resulting clusters is enhanced. However, as the cluster size become higher than 4, $C$ starts to increase. This indicates that by obtaining more than 4 clusters, some of these clusters will have correlated information, and thus are better to be merged to form larger clusters. As $C$ has the lowest value for cluster size of 4, the optimal number of cluster for this dataset is obtained as 4.

TABLE II
SIGNIFICANT FEATURES TIME PERIOD

| Time Instances | | |
|---|---|---|
| 16-Feb-2016 13:00:00 | 24-May-2015 06:00:00 | 17-Jan-2016 11:30:00 |
| 30-Jul-2015 23:30:00 | 31-Jul-2015 21:00:00 | 12-Feb-2016 19:00:00 |
| 15-Apr-2015 04:00:00 | 08-Apr-2015 10:30:00 | 23-Dec-2015 04:30:00 |
| 15-May-2015 10:30:00 | 04-Aug-2015 12:30:00 | 11-Mar-2015 18:00:00 |

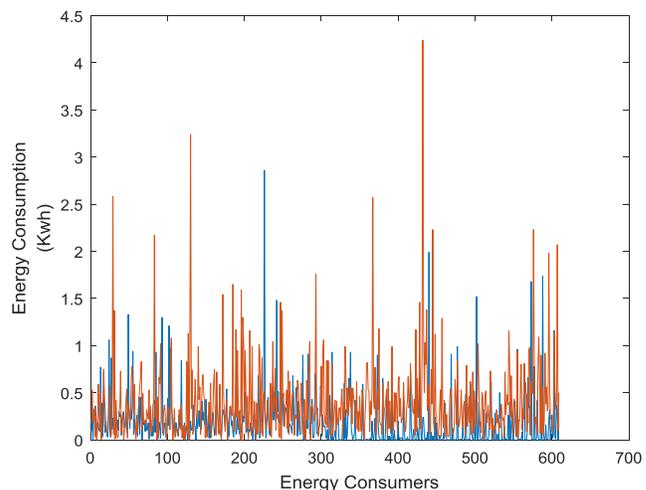

Fig. 4. Selected features at two time instances.

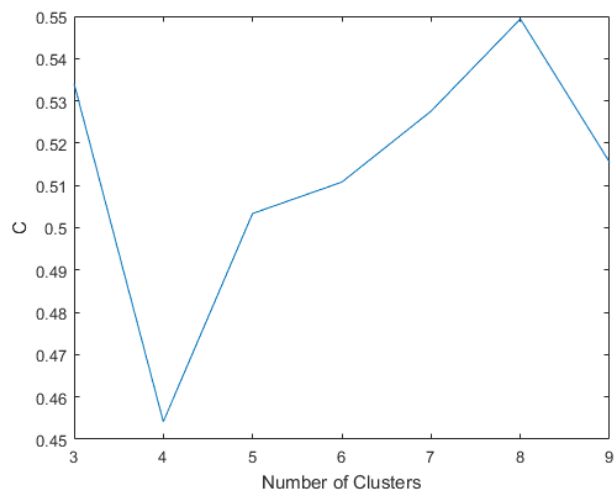

Fig. 5. The proposed metric $C$ as a function of the number of clusters. $C$ measures distinguishability of clusters and the optimal cluster size is the case with minimum $C$.

Analyzing characteristics of each cluster of energy consumption profile provides knowledge discovery in terms of the time instance in which the maximum demand on the network is recorded, and hence efficient future planning can be achieved. According to a report published by the distribution company of which the data is obtained, "Peak demand typically occurs on five to eight days of the year, and for around five hours in total on each of those days" [25]. Looking at Figure 6, which shows the energy consumptions of the obtained optimal clusters, one can find the maximum energy consumption and hence the demand peaks in two time instances; one in July and one in December. This shows that future planning can include demand response programs to manage the peaks on the power network. The clusters' consumption profile also provides information on which group to target for demand response. For example, to manage the first peak in July, cluster 3 will likely provide the best outcome as it has 30% of the energy consumers falling in that cluster as shown in Figure 7. However, to manage the second peak in December, cluster 1 will likely provide the best outcome even though it comprises of only 17% of the energy consumers (Figure 7). A combination of more than one cluster can be targeted for demand response to achieve further reduction. As demand side reposes and finding the optimal demand side management policy is not in the scope of our work in this manuscript, we will further work on that in the future. We have indeed an ongoing collaboration work with the distribution industry on effective demand side response, and the outcome will be published in future works.

5, and 8, respectively. While both peaks are clear in Figure 8, clusters 1 and 3 have similar pattern during the first peak with combined percentage of energy consumers of 66% as shown in Figure 11. This limits the number of options for targeting consumers to manage the first peak. The same pattern is observed in the second peak where the combined percentage of energy consumers is 53%. One distinct difference between this case and the one with the optimal cluster size is observed in the period between June to December. The case with the optimal cluster size shows that among two profiles, one of them is higher than the other one between May to September, which then flips in the period between September to December. This detail is lost when the cluster size is set at 3, where these two clusters are combined into one.

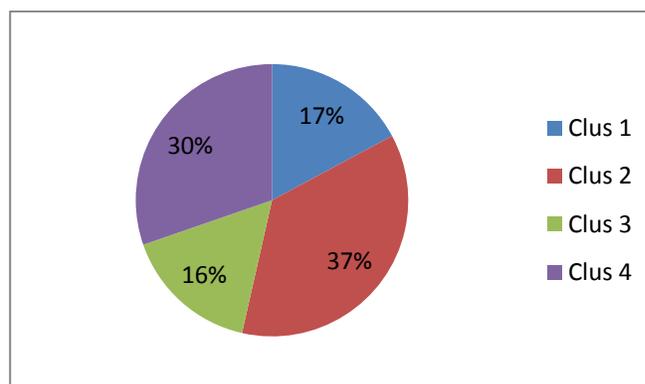

Fig. 7. Percentage of energy consumers in each cluster when $C$ is 4. Cluster 2 has the highest percentage of energy consumers followed by cluster 4.

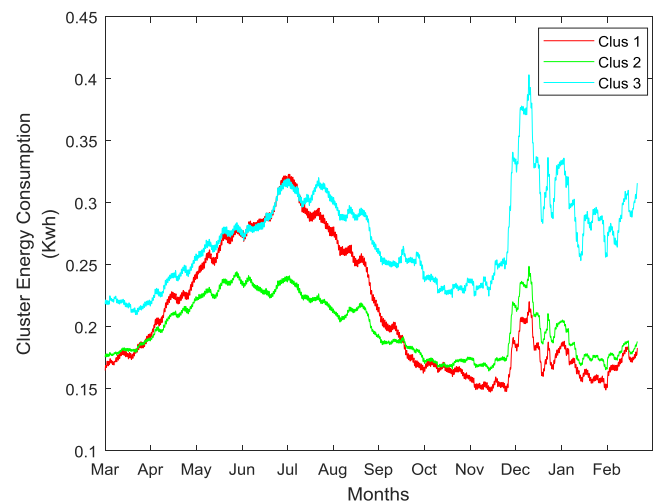

Fig. 8. Average energy consumption of clusters when the cluster size is set at 3.

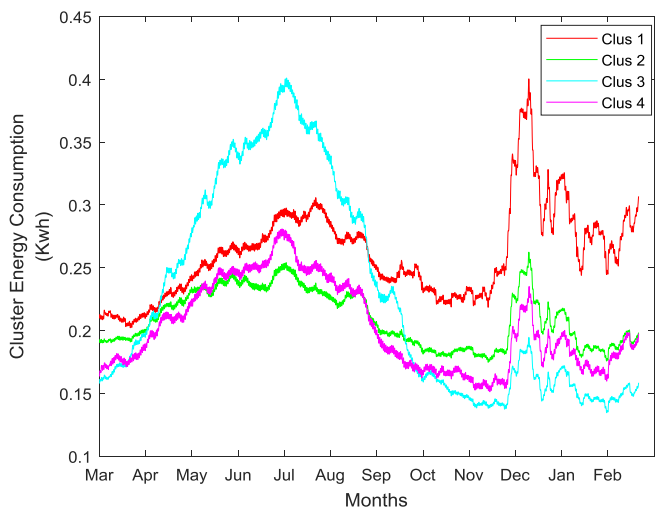

Fig. 6. Average energy consumption of optimal clusters across the year.

To understand the importance of optimal cluster size in terms of knowledge discovery and other benefits, further analysis is required to study the role of non-optimal cluster size on the performance. Here, we further analyze clusters of three, five and eight of the 609 energy consumers dataset, which all have higher $C$ than the case with cluster size 4, and thus are not optimal. Figures 8, 9 and 10 depict the clusters' average energy consumption when the cluster size is set at 3,

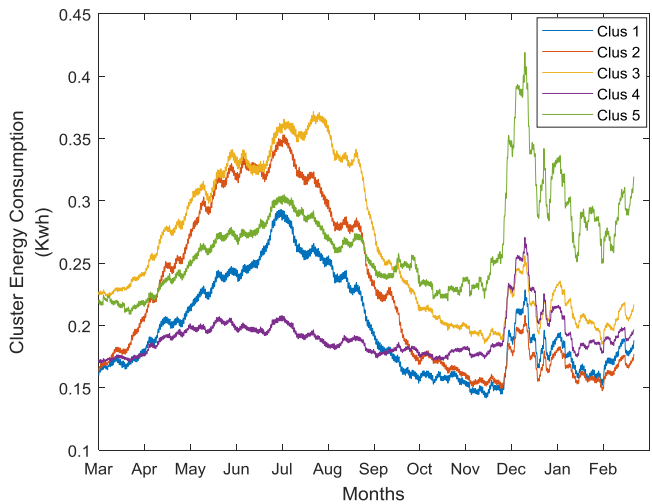

Fig. 9. Average energy consumption of clusters when the cluster size is set at 5

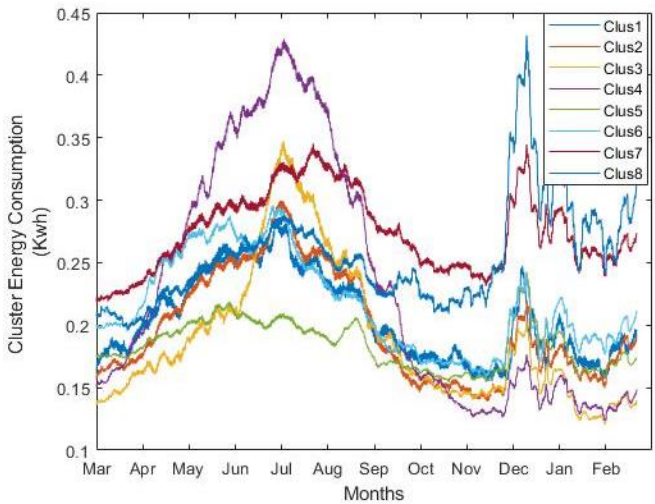

Fig. 10. Average energy consumption of clusters when the cluster size is set at 8.

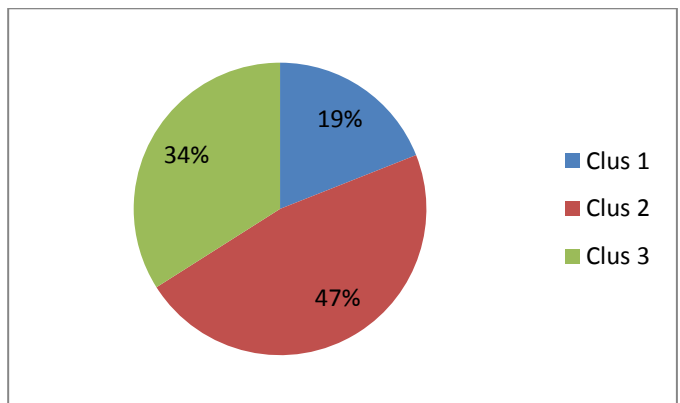

Fig. 11. Percentage of energy consumers in each cluster when $C$ is 3. Cluster 2 has the highest percentage of energy consumers whereas cluster 1 has the lowest.

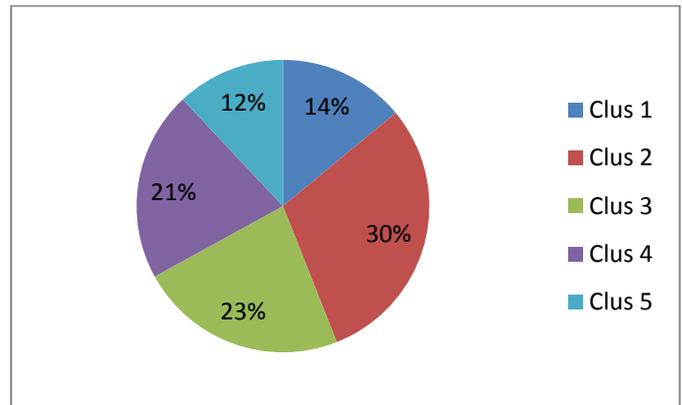

Fig. 12. Percentage of energy consumers in each cluster when $C$ is 5. Cluster 2 has the highest percentage of energy consumers whereas cluster 5 has the lowest.

Figure 9 introduces further cluster profile that depicts similar pattern compared to the optimal clusters, but being different in intensities split into two new clusters. It can be observed from Figure 12 that the percentage of energy consumers is further decreased in this case. It is also worth noting that 5 clusters is considered a suboptimal clustering for the dataset and hence can be used for various programs including customized pricing policies and demand response. Figure 10 dilutes the clusters even further and creates similar clusters with lower percentage of energy consumers compared to the optimal and suboptimal clusters as shown in figure 13. This creates inefficiency as SOM uses additional computational resources to cluster the dataset into further smaller clusters. It is also not a straight forward decision as to which clusters to target as many of the clusters have similar profile across the year.

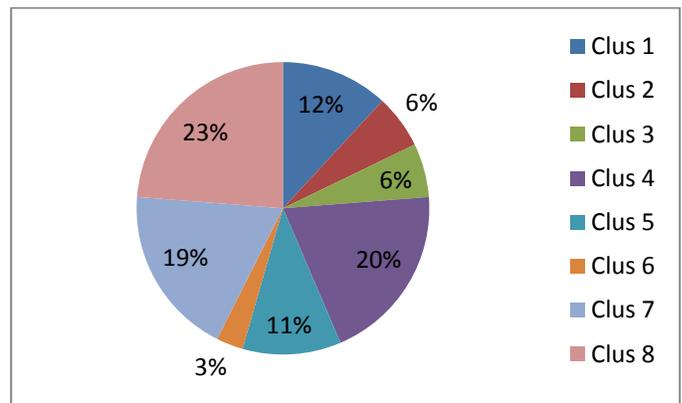

Fig. 13. Percentage of energy consumers in each cluster when $C$ is 8. Cluster 8 has the highest percentage of energy consumers whereas cluster 6 has the lowest.

*B. Dataset 2*

Using the same approach utilized on dataset 1, the GA-based feature selection was applied to dataset 2 to find the most significant features in the dataset. The dataset is then clustered into groups using SOM and the proposed metric $C$ was used to find the optimal number of clusters. As depicted in Figure 14, the optimal number of clusters is found to be 15 where the case with 8 clusters also resulted in close performance to the optimal case. As it can be seen from Figure 15, the number of energy consumers in clusters 13, 14 and 15 is relatively the highest compared to other clusters. It can be interpreted from Figure 18 that the average annual energy consumption of clusters 13,14 and 15 is 0.66, 0.44 and 0.25 kWh respectively. These households are considered low consumers in the dataset compared with others. Similarly, the number of energy consumers in clusters 7, 9 and 12 are close to each other. It can be interpreted from Figure 17 that the annual energy consumption of clusters 7, 9 and 12 is 1.1, 0.7 and 2.1 kWh respectively. These households are considered moderate consumers in the dataset compared with others. On the other hand, the number of energy consumers in clusters 1, 2, 3, 4, 5, 6, 8, 10, 11 is the lowest in the dataset. It can be interpreted from Figure 16 that the annual energy consumption of clusters 1, 2, 3, 4, 5, 6, 8, 10, 11 is 16.6, 9.5, 10, 7.6, 3.6, 3.2, 1.1, 4.8 and 4.3 kWh respectively. These households are considered high energy consumers in the dataset compared with others.

From the analysis of the Irish dataset, the demand side response can be designed to suit three main categories; low, medium and high energy consumers.

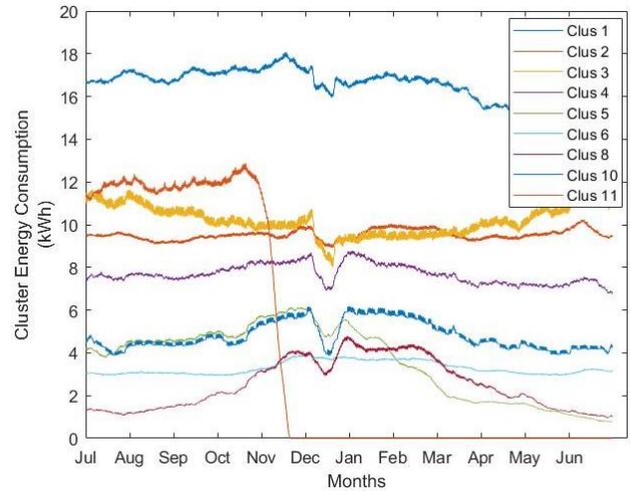

Fig. 16. Average energy consumption of clusters 1,2 , 3, 4, 5, 6, 8, 10 and 11 from the optimal number of clustering. Energy consumption ranges from 0 kWh for cluster 8 to 18 kWh for cluster 1

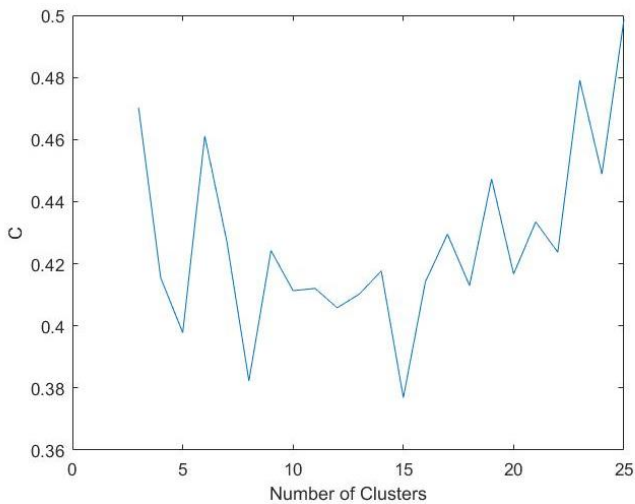

Fig. 14. The proposed metric $C$ as a function of the number of clusters. $C$ measures distinguishability of clusters and the optimal cluster size is the case with minimum $C$ which is 15 clusters.

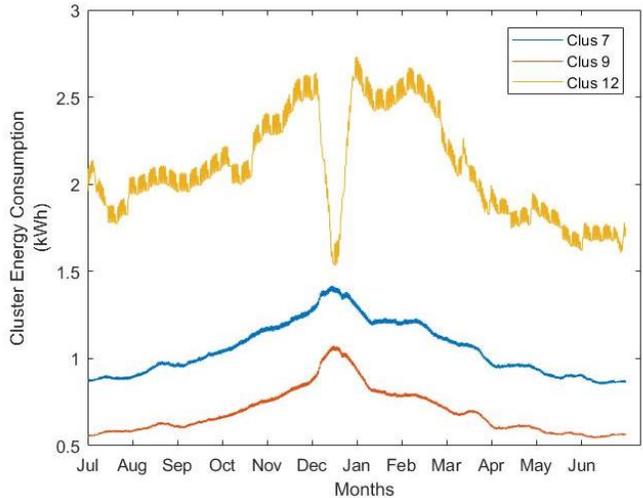

Fig. 17. Average energy consumption of clusters 7, 9 and 12 from the optimal number of clustering. Energy consumption ranges from 0.0.5kWh for cluster 9 to 2.6 kWh for cluster 12

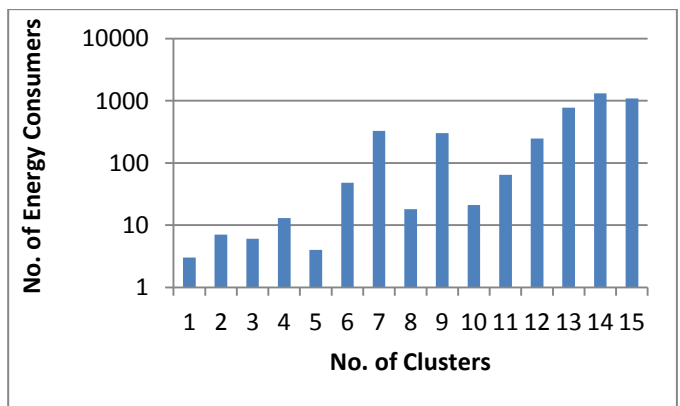

Fig. 15. Number of energy consumers in each cluster from the Irish dataset. Majority of consumers are in clusters 13, 14 and 15.

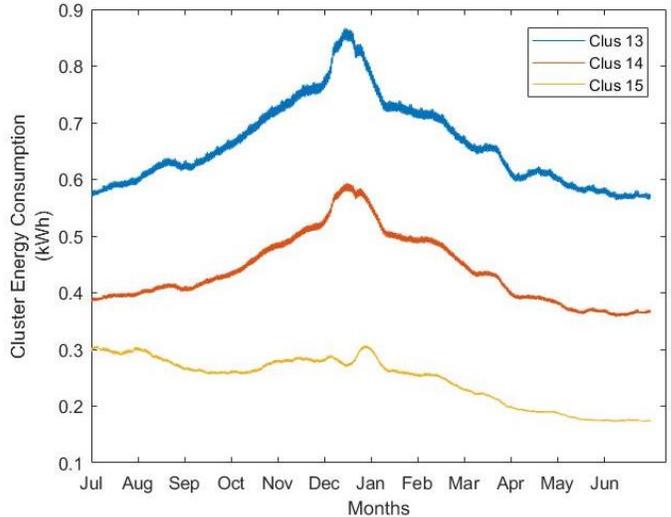

Fig. 18. Average energy consumption of clusters 13, 14 and 15 from the optimal number of clustering. Energy consumption ranges from 0.2kWh for cluster 15 to 0.88 kWh for cluster 13

## V. Conclusion

A necessary step before demand side management policies is to identify customers' segments and group them into distinct clusters based on their energy consumption rates. An important step in any clustering algorithm is to find the optimal cluster size. This manuscript presented an optimal clustering metric for grouping energy consumers into clusters. While clustering techniques based on energy consumption exists in the literature, the concept of optimal clustering of a dataset is new. It has been shown that optimal clustering brings many benefits including the identification of groups for targeted programs such as demand response programs or customized energy pricing. We proposed a metric based on the entropy of the eigenvalues of the correlation matrix between the clusters. The proposed metric measures distinguishability between the clusters, and its minimum indicated the optimal cluster size. We also used a feature selection method based on genetic algorithms to obtain an optimal set of feature for the clustering. Self organized maps were used for the clustering task. Our numerical results on two sample datasets, one obtained by an energy distribution company located in Victoria, Australia and one obtained by Irish Commission for Energy Regulation, showed that one can successfully find the optimal cluster size for the dataset. We further showed that the clusters with the optimal cluster size have indeed the optimal information for decision making purpose, as compared to those with non-optimal cluster size. Our future work is to set an optimal demand side response based on the optimal clusters obtained by this method.